# Linear Unitary Cellular Automata and Convolution Algebra

Theophanes E. Raptis[1]

[1] Computational Applications Group, Division of Applied Technologies, National Centre for Science and Research "Demokritos", 153 10, Aghia Paraskevi, Athens, Greece

Corresponding author: Theophanes E. Raptis (e-mail: rtheo@dat.demokritos.gr)

**Abstract**:

The original local, discrete example of Linear Unitary Cellular Automata (LUCA) is analyzed in terms of a new representation previously introduced in [1] for classical CA. Several important underlying symmetries are reviewed and their tight relationship with both signal and coding theory as well as with combinatorics is underlined. A class of analog implementations in the form of Linear Transmission Line Networks (LTLN) is described as possible emulators of this type of dynamics.

Keywords:

*Cellular Automata, Unitary Evolution, Convolution, Transmission Lines*

1. **Introduction**

Recently, a new interpretation of Cellular Automata (CA) dynamics was introduced in [1] where any n-dimensional CA is decomposed in a product of two maps as $(R.f)(l)$ over a lattice $l$ where $l$ is a one dimensional reshape of the original n-D lattice, $f$ is a linear map $f:\{0,\ldots,b\} \rightarrow \{0,\ldots,b^{|N|}\text{-}1\}$: $f = \hat{C} \cdot l$ where $|N|$ is the number of neighbors in an arbitrary shape neighborhood, $C$ is a Circulant Convolution matrix and $R$: $\{0,\ldots,b^{|N|}\text{-}1\} \rightarrow \{0,\ldots,b\}$: $l_{n+1} = R(f_n) = R(\hat{C} \cdot l_n)$ is the non-linear part. The above definitions will hold true for any neighborhood topology of any lattice taking values from an arbitrary symbolic alphabet in base $b$. Any topological information and resulting correlations are preserved by the structure of the circulant matrix thus representing an ultimate dimensional reduction scheme. This is only possible here due to space-time discreteness which makes dimensionality trivial.

The introduction of circulant matrices allows using digital and analog signal processing techniques due to the particularly nice property of any such matrix which gets diagonalized by DFT matrices as $C = F^{-1} \cdot \Lambda \cdot F$. This allows examination of the overall dynamics in the dual space $\tilde{h} = \Lambda \cdot \tilde{R}(h)$ where bars denote the DFT of each vector. Furthermore, the possibility of a direct reversible analog implementation has been examined via a direct transfer of the previous scheme in the frequency or wavelength domain via an extension of the already known "Manchester" encoding [2] into a class of permutation machines in the Fourier domain which by construction has a unitary structure.

Almost two decades ago, different classes of Linear Unitary CA (LUCA) were introduced by Grössing and Zeilinger [3] [4] as well as Bialinycki-Birula [5], [6] and which can also be brought in the same format as with the previous fully nonlinear classical CA.. In fact, the complete absence of the nonlinear $R$ ("rule") map in the case of Birula and other such UCA makes this case attractive for studying them as special cases of signal theory. To make this visible requires a rearrangement of the elements originally defining the lattice interactions which were given in the form of a "scattering" matrix between states of different neighbors. This is examined by an extension of the previously introduced linear map $f$: $C^2 \rightarrow C^2$ to include local values of a spinorial wavefunction. We restrict attention to the case of 1D UCA as well as that of Birula class simulating Weyl particles, the extension to the other two cases being straightforward.

The reason for making a strong discrimination between linear UCA and others came from recent advances in the field where certain nonlinear extensions were added by Elze [37] and others in the context of so called emergent quantum mechanics [38] which are beyond the scope of the present work and which is more closely associated with alternative implementations with actual electronic and/or electro-optical technology based on an appropriate reformulation of their dynamics. Yet, the previous authors too have given emphasis on the inherent links of signal and information theory with quantum automata albeit in a different context. Hopefully, the present work extends this line of thought at least in the standard context.

In the next section, we make a preliminary examination of certain important properties of circulant convolution matrices that are generic to all linear automata that will be examined using a weaker one-dimensional (1D) case introduced by Grössing at the 80s. In section 3, the full 3D case of Birula complete UCA is further analyzed in terms of certain internal symmetries found, while in section 4, a Kronecker factorization of the resulting kernel is proposed that can be expanded in arbitrary dimensions. All numeric and symbolic codes used for evaluation of matrix operations and graphics are currently available through the author's github account [39]. In section 5, a new analog machine design is proposed which can be used to reduce the information of a multi-dimensional lattice into a set of aperiodic signals via Fourier encoding in frequency or wavelength domain.

## 2. A simple one 1D model

In an early attempt to introduce the CA paradigm in quantum lattices via the Heisenberg representation, Grosser and Zeilinger offered an approximate complex valued CA version of a 1D tight-binding Hamiltonian with nearest neighbor interaction. They then use a $1^{st}$ order approximation to derive an almost unitary evolution for small values of the expansion parameter. We can find an exact solution for such a model based on the fact that any such Hamiltonian is by construction circulant.

$$H = \begin{bmatrix} 0 & \bar{\varepsilon} & 0 & \ldots & \varepsilon \\ \varepsilon & 0 & \bar{\varepsilon} & 0 & \ldots \\ & \ddots & \ddots & \ddots & \\ & & \ddots & \ddots & \varepsilon \\ \bar{\varepsilon} & \ldots & 0 & \varepsilon & 0 \end{bmatrix}$$

The fundamental properties of such matrices are well known in signal theory and the most important are known to be their special diagonalization via DFT matrices [7]. It is then immediately diagonalizable via

$$H = F^{-1} \Lambda_\varepsilon F, \quad \Lambda_\varepsilon = F \cdot \mathbf{h}_1 \qquad (1)$$

The vector $\mathbf{h}_1$ in (1) stands for the $1^{st}$ row of $H$, $[0, \bar{\varepsilon}, 0, \ldots, \varepsilon]$ of which the DFT provides all eigenvalues immediately. From the representation of any DFT matrix as a Vandermode matrix on the $L$th roots of unity of a length $L$ lattice, and using the original choice for the matrix elements as $\varepsilon = \varepsilon_0 (1 + \mathbf{i})$ one obtains the eigenvalue spectrum as

$$\lambda_i = \frac{\varepsilon_0}{\sqrt{L}}(1+\mathbf{i})\omega^i + \frac{\varepsilon_0}{\sqrt{L}}(1-\mathbf{i})\omega^{i(L-1)} = \frac{2\varepsilon_0}{\sqrt{L}}(\cos(w) - \sin(w)) \qquad (2)$$

In (2), we have taken $w = 2\pi i / L$ and we used the internal symmetry of the DFT matrix under which the two different powers of the upper and lower triangular parts across the same row are complex conjugates and $w = 2\pi i / L$. Moreover, for the propagator we can immediately derive its circulant form via the map

$$\exp\left[-\mathbf{i}\left(F^{-1}\Lambda_\varepsilon F\right)t/\hbar\right] \rightarrow F^{-1}\left(e^{-\mathbf{i}\Lambda_\varepsilon t/\hbar}\right)F \qquad (3)$$

Given an arbitrary initial condition $y = \sum_{i=1} c_i |i>$, the evolution can be written directly in a plane wave basis as

$$\tilde{y}(k)_{t+\delta t} = e^{-\mathbf{i}\Lambda_\varepsilon(k)t/\hbar}\tilde{y}(k)_t, \quad \tilde{y} = F \cdot [c_1,...,c_n]^T \qquad (4)$$

We immediately notice that taking the diagonal representation is identical with the algebraic Hadamard (componentwise) product with the vector given by $Diag[\exp(-\mathbf{i}\Lambda_\varepsilon(k)t)]$. This can then be put into a formal correspondence with an inverted convolution product. Using the general convolution identity $F(G)F(y) = F(G \otimes y)$ with $F(G) = \exp(\mathbf{i}\lambda t/h)$ we deduce that the action of a circulant Hamiltonian can be recast as a form of spectral evolution of the cross-correlation of the initial condition and the Fourier inverse of the propagator in the special form

$$y(x,t) = \left(F^{-1}e^{\mathbf{i}\lambda(k)t/\hbar}F\right)(G \otimes y_0)$$
$$G \otimes y_0 = \int dx' G(x-x',t)y(x',t=0) \qquad (5\text{a-c})$$
$$G(x,t) = F^{-1}\left(\exp(-\mathbf{i}\lambda(k)t/\hbar)\right)$$

In (5b), we can always interpret the convolution as the cross-correlation of $G(-x,t)$ which effectively gives the evolution probability density of $|G(t)|^2 + |y_0|^2$ in time. Unitarity then demands that $|G|^2$ remains constant as a product of two unitary operations. One may utilize the above observation for finding more general kernels $G$ appropriate for unitary evolution.

Two more general observations are due at this point, regarding certain evolutions in algorithmic structures relevant for such problems of which the physicist's community seems unbeknownst at large. The possibility of expressing arbitrary matrices via products of diagonal and circulant matrices [8], [9] a fact that is tightly connected with applications in diffractive optics [10] opens a way towards efficient, exact diagonalization algorithms which could be transcribed in analog, optical machines.

While the methods mentioned in the literature are quite general and could include even random matrices, there is at least one important case of intermediate complexity which can be used as an example in the area of Bose-Hubbard Hamiltonians [11] [12]. This can be understood via a direct map of any Fock representation of many particles given as $|n_1, n_2, ..., n_L>$ into the lexicographically ordered powerset of the integers expressed in a higher alphabet of base $b = \max(\{n_i\}_{i=1}^L)$. This property is revisited in section 5 and in the relevant App. C in association with Kronecker product structures. Given a particle conservation constraint as $N = \sum_{i=1}^{L} n_i$ the problem of identifying all Fock vectors with constant population becomes identical with the combinatoric problem of finding all restricted integer partitions of $N$ [13], [14]. This set has recently been found to contain a self-similar structure by Folsom, Kent and Ono [15]. The particular type of primitive arithmetic self-similarity can be traced back at the nature of all lexicographically ordered powersets of high alphabet words being identical with a discrete sampling of a product of harmonic oscillators of exponential ("lacunary") scaling, the simplest one given by the so called, Rademacher basis [16] for the binary case. Algorithms extracting all such partitions solely from the recursive block structure of the binary system exist but they lie beyond our present scope and they will be reported elsewhere.

Regarding the Bose-Hubbard model, given an ordered set of Fock vectors, the action of any such Hamiltonian as well as others can be reduced into a mere string matching automaton. That is, given any abstract 2$^{nd}$ order quantized operator pair $a_{i+k} a_{j+l}^+, 0 \leq k, l \leq b-1$, its sole action is to shift the values of individual symbols from a word on a $Z_b$ ring. Given that any such ordered set of words can be represented as an $N \times b^N$ array of $b$ "colors", it is relatively easy to extract the appropriate matrix elements as a Cartesian product of two such arrays. This then resembles the action of a particular Turing automaton moving across the large axis of the ordered array according to the information contained in the operator indices. Via a direct application of the polynomial representation one can extract a matching condition between matrix indices of an arbitrary abstract operator pair in the arithmetized form

$$|n_1, n_2, ..., n_L> \cong n \in \left[0, ..., b^L - 1\right] : (a_{i+k} a_{j+l}^+)(n) \to n + b^{j+l} - b^{i+k} \qquad (6)$$

The significance of ordered powersets seems to have been recognized in a recent publication by Zhang on exact diagonalization techniques [17] yet it seemingly remains unknown by the community at large as can be seen in other more recent attempts [18] that use more primitive toolboxes for similar purposes. The above serve as s first step to revealing the strong inherent ties between physical processes and seemingly different areas like number theory, combinatorics, general Automata theory and computational structures. The isomorphism between certain Hamiltonian structures and automata deserves further examination which will be the subject of a future work.

While in the 1D case, derivations based on circulant symmetry of the interaction kernel are relatively easy, situation is more difficult with inclusion of spinorial structures on a full 3D lattice. Fortunately, a complete treatment of this issue for various types of both integer and half-integer spin has been provided by Bialinycki-Birula and its ramifications are explored in the next section.

## 3. Reformulation of the 3D Bialinycki-Birula LUCA

Following the formulation given by Birula, the particular type of 3D lattice is first given in a form identical with that of CsCl cubic lattice of 8 neighbors and with lattice sites occupied by spinorial wavefunctions $y(r, t) = (\varphi(r, t), \xi(r, t))$ at $t = 0$ evolving with discrete time steps at $t + \delta t$. This choice of a lattice has been made by Birula in order to comply with certain constraints required for both unitarity as well as the Lie-Trotter approximation used in the limit [4]. This does not necessarily exhausts all possible types of automata that could perhaps exhibit similar behavior.

In the original work, the UCA update rule is described through a set of 8, 2 x 2 complex matrices $W$ so as to have

$$y(\mathbf{r}, t + \delta t) = \sum_{\mathbf{h}} \mathbf{W}_{\mathbf{h}} y(\mathbf{r} + \mathbf{h}, t) \tag{7}$$

In (7), the choice of neighboring lattice vectors is numbered by a convention as the set of the 8 bit strings in the alphabet {-1, +1} from {-1, -1, -1} up to {1, 1, 1} with $2^3$ combinations in total. To ease the presentation we will have to change this into standard binary notation so that in this presentation $y_1$ will stand for {-1, -1, -1} and $y_8$ for {1, 1, 1} while 0 index is preserved for the central 9$^{th}$ cell.

The $W$ matrices satisfy certain criteria in order to have a proper resolution of the identity and completeness in the corresponding Hilbert space. Their exact form can be found in the original papers and instead of reproducing them here we will directly present the result of applying them in a set of neighboring wavefunctions. Using symbolic software is fairly easy to check that the expansion of (1) with the matrix elements originally defined gives

$$\phi_0^{t+\delta t} = \frac{\alpha}{4}(\xi_2 - \xi_3 + \phi_5 + \phi_8) + \frac{\beta}{4}(-\xi_1 + \xi_4 + \phi_6 + \phi_7)$$

$$\xi_0^{t+\delta t} = \frac{\alpha}{4}(\xi_2 + \xi_3 - \phi_5 + \phi_8) + \frac{\beta}{4}(\xi_1 + \xi_4 + \phi_6 - \phi_7) \tag{8}$$

In (8) we use the parametrization $\alpha = 1 + i$ and $\beta = 1 - i$. Eq. (8) hides an intrinsic symmetry which can be made more evident by further symmetrization of the update operation.

During each update only half of the upper spin parts are utilized for the first four neighbors as well as half of the lower spin parts for the last four. Indeed, we find that it is possible to expand the output of (8) as a 4 vector in the form $\bar{\Psi} = [\phi, \xi, \xi^*, \phi^*]$ where (*) denotes complex conjugation and the two input vectors are of the form $\vec{\Phi} = [\phi_5, \phi_6, \phi_7, \phi_8], \vec{\Xi} = [\xi_1, \xi_2, \xi_3, \xi_4]$. This way we obtain a full 4 x 4 matrix update equation as

$$\vec{\Psi} = \mathbf{K} \cdot \vec{\Phi} + (\mathbf{P} \cdot \mathbf{K}) \cdot \vec{\Xi} \qquad (9)$$

In (9), the two new matrices are of the form

$$\mathbf{K} = (1/4) \begin{bmatrix} \alpha & \beta & \beta & \alpha \\ -\alpha & \beta & -\beta & \alpha \\ -\beta & \alpha & -\alpha & \beta \\ \beta & \alpha & \alpha & \beta \end{bmatrix}, \quad \mathbf{P} = \begin{bmatrix} 0 & 0 & 1 & 0 \\ 0 & 0 & 0 & 1 \\ 1 & 0 & 0 & 0 \\ 0 & 1 & 0 & 0 \end{bmatrix}$$

The construct in (9) could be interpreted as the simultaneous co-evolution of two dual lattices the second under spin exchange and conjugation if (9) was to be taken with its conjugate to form a system of two automata, the second living on a lattice shifted by one position inside a larger augmented 27-cell neighborhood of a normal cubic lattice. We also observe that the *K* matrix has an interesting property of being "anti-Hadamard" satisfying the relations

$$\mathbf{K} \cdot \mathbf{K}^T = (1/2)\mathbf{J}, \quad \mathbf{K}^T \cdot \mathbf{K} = \mathbf{P} \qquad (10)$$

where *J* is the 4 x 4 exchange matrix with ones only in the anti-diagonal. This is solely due to the particular arrangement of neighbors which causes a permutation of the basis of orthogonal rows.

Inspired by this observation, we seek after a way to reformulate (9) in terms of a pure Hadamard transform. Indeed this is possible by separation of real and imaginary parts in (8) in which case we obtain a new update rule this time for the symmetrized vector $\vec{\Psi}' = [\phi, \xi, \xi, \phi]$ as

$$\vec{\Psi} = (1/4)\mathbf{K}_1 \cdot \vec{\Phi} + (1/4)\mathbf{K}_2 \cdot \vec{\Xi}$$

$$\mathbf{K}_1 = \mathbf{P}_1 \cdot \mathbf{H}_1, \quad \mathbf{K}_2 = \mathbf{P}_2 \cdot \mathbf{H}_2 \qquad (11)$$

In (11), the two Hadamard matrices $\mathbf{H}_i$ are obtained in the form

$$\mathbf{H}_1 = \begin{bmatrix} 1 & 1 & 1 & 1 \\ 1 & -1 & -1 & 1 \\ -1 & 1 & -1 & 1 \\ -1 & -1 & 1 & 1 \end{bmatrix}, \quad \mathbf{H}_2 = \begin{bmatrix} 1 & 1 & 1 & 1 \\ -1 & 1 & 1 & -1 \\ -1 & 1 & -1 & 1 \\ 1 & 1 & -1 & -1 \end{bmatrix}$$

Their accompanying connectivity matrices **P** that restore the original wavefunctions from their corresponding real and imaginary parts are also of the form

$$\mathbf{P}_1 = \begin{bmatrix} 1 & i & 0 & 0 \\ 0 & 0 & 1 & i \\ 0 & 0 & 1 & i \\ 1 & i & 0 & 0 \end{bmatrix}, \quad \mathbf{P}_2 = \begin{bmatrix} 0 & 0 & 1 & i \\ 1 & i & 0 & 0 \\ 1 & i & 0 & 0 \\ 0 & 0 & 1 & i \end{bmatrix}$$

Having completed the analysis of the update rules, we may proceed to a reconstruction of the dynamics in the reduced dimensionality picture via the introduction of an appropriate Circulant matrix. The above analysis indicates the presence of some variant of a wavelet basis, a theme which is further discussed in the following and the last section.

4. **Equivalent representations with circulant convolution kernels**

For an arbitrary multi-dimensional automaton, unfolding of a hypercubic lattice into a 1D configuration requires a protocol for transferring the neighbor correlations into a generic circulant interaction kernel. As any such kernel is constructed by the elements of its first row, it suffices to isolate matrix elements from the update rules that preserve the connectivity in the 1D projection of an original n-D lattice. The n-D lattice is then read in the same manner that multi-dimensional arrays are realized in modern computers via an appropriate pointer setting algorithm. This algorithm has as a peculiar side effect the possibility of a proof for the existence of infinite disordered arithmetic bases of which a sketch is given in App. A.

For the original Cs-Cl lattice proposed by Birula, we face a particular problem with reduced symmetry in that any direct unfolding of an augmented 27 cell neighborhood would lead to alternating between odd-even addresses. To avoid this one needs a careful dissection of the original lattice to two separate cubic lattices $\{L_1, L_2\}$ of $N^3$ sites such that any central cell of the original lattice is contained in $L_2$ with its neighbors in $L_1$. To properly address issues with boundary conditions we take all interior positions corresponding to $L_2$ to be equal to the exterior positions of $L_1$ shifting across all sites of $L_2$ by one. Hence if the original contained $N^2$ cells at the three external boundary faces for $L_1$ corresponding to the planes $\{1,1,0\}$, $\{1,0,1\}$, $\{0,1,1\}$ then $L_2$ will contain the same amount of cells with all edge positions moved at $N+1$ at opposite faces. To complete the reduction we also need to separate the original set of spinorial wavefunctions in two sets for each lattice.

Following the logic of the update equation (8), we see that updated values alternate between the two new lattices as each cell becomes the central cell for the other. Then the two update steps can be executed in tandem via the equations

$$\Phi^{(1)}{}_{n+1} = C_1 \Phi^{(2)}{}_n + C_2 \Xi^{(2)}{}_n$$
$$\Xi^{(1)}{}_{n+1} = C_3 \Phi^{(2)}{}_n + C_4 \Xi^{(2)}{}_n \tag{12a}$$

$$\Phi^{(2)}{}_{n+1} = C_1 \Phi^{(1)}{}_n + C_2 \Xi^{(1)}{}_n$$
$$\Xi^{(2)}{}_{n+1} = C_3 \Phi^{(1)}{}_n + C_4 \Xi^{(1)}{}_n \qquad (12b)$$

In this case, we have to introduce four circulant kernels of which the construction is analyzed in more detail in App. B where a proof of a kronecker factorization property for such kernels is given. What becomes immediately apparent in (12a-b) is the possibility of total decoupling of the two lattices for fixed boundary conditions by folding the two operations for every second time step as

$$\Phi^{(i)}{}_{n+2} = C'_1 \Phi^{(i)}{}_n + C'_2 \Xi^{(i)}{}_n$$
$$\Xi^{(i)}{}_{n+2} = C'_2 \Phi^{(i)}{}_n + C'_3 \Xi^{(i)}{}_n \qquad (13a)$$

$$C'_1 = C_1^2 + C_3^2, \quad C'_2 = C_3 \cdot (C_1 + C_4), \quad C'_3 = C_4^2 + C_3^2 \qquad (13b)$$

Application of (13a) presupposes that two separate initial conditions have been prepared of which the second is the image of the first under the original evolution of (12a-b) so that they evolve independently for even and odd time steps respectively. When the $C_i$ kernels are perfectly circulant, eigenvalues are directly obtainable from the DFT of the first row of each one. One also has the obvious identities

$$C_i \cdot C_j = (F^{-1} \Lambda_i F)(F^{-1} \Lambda_j F) \cong F^{-1} \Lambda_{ij} F, \quad \Lambda_{ij} = \Lambda_i \cdot \Lambda_j .$$

Additionally, the summand of two circulants also remains circulant. While this holds for fixed B. C., it breaks down under an additional circular shift needed to account for edge cells in periodic conditions taken (*mod N*) in the original 27-cell augmented neighborhood. In such a case, each sub-block in the kernel's Kronecker factorization remains almost circulant but a defect is introduced at the edges of each block.

The case of Block Circulant matrices is also known and given each block's corresponding DFT diagonalizing matrix, the total eigenvector matrix can be given via the Kronecker product of their lower dimensional component DFT sub-matrices. A general treatment of block circulant matrices can be found in [19]. In App. B, we provide guidelines on how to construct the kernel matrices as well as for their simplification via Kronecker factorization. Several references on factorization methods for Kronecker products with signal processing applications can be found in [20]-[23]. Using the total eigenvalues from (13b) one gets an equivalent diagonal representation of (13) as

$$\tilde{\Phi}^{(i)}{}_{n+2} = \Lambda_1 \tilde{\Phi}^{(i)}{}_n + \Lambda_2 \tilde{\Xi}^{(i)}{}_n$$
$$\tilde{\Xi}^{(i)}{}_{n+2} = \Lambda_2 \tilde{\Phi}^{(i)}{}_n + \Lambda_3 \tilde{\Xi}^{(i)}{}_n \qquad (14)$$

In figure 1a, a plot of the amplitudes of the eigenvalue spectra computed as one of the four circulant kernels is shown while in figure 1b, the same is shown for the three composite kernels of the decoupled equations (14). At this point, we notice the additional significance of the characteristic rows in that they form an extended basis similar to that of the original matrices in (9). Indeed, given their orthogonality property as mentioned in App. B, they serve as a kind of wavelets in that if taken also as initial conditions the action of the circulant kernels results in a simple power of the corresponding eigenvalue which is then equivalent on a shift of the original preimage. Hence, motion is represented this way for individual "wavicles" while scattering corresponds to the mixing of powers of different eigenvalues.

In the final of the evolution equations (14) we also have a total decoupling so that they can be represented separately for each cell via the transfer matrix $M$ and its relevant SU(2) decomposition in the Pauli basis given from the traces of $\mathbf{M} \cdot \sigma_i$ as

$$\hat{\mathbf{M}}_k = \begin{pmatrix} \lambda_{1,k} & \lambda_{2,k} \\ \lambda_{2,k} & \lambda_{3,k} \end{pmatrix} = \frac{\lambda_+}{2} (I + \lambda_x \sigma_x + \lambda_z \sigma_z)$$

$$\lambda_+ = \lambda_{1,k} + \lambda_{3,k}, \lambda_x = \lambda_{2,k} / \lambda_+, \lambda_z = (\lambda_{1,k} - \lambda_{3,k}) / \lambda_+$$

(15)

The same exactly can of course be derived from the diagonal representation of the original one step coupled equations 12(a-b) with a simpler set of local eigenvalues. The long term evolution of any cell can then be described for the powers of the noncommuting trinomial in the *lhs* of (15). Notably, any such multinomial expression can be reduced to a superposition of all words in the ordered powerset of strings of increasing length from a ternary alphabet via the correspondence $\mathbf{I} \to '0', \sigma_x \to '1', \sigma_z \to '2'$ where each concatenated set of symbols is then interpreted as a partial product of order *n* giving rise to higher order polynomials of the eigenvalues. Apparently, the set of sequences collected this way contains the natural periodicities from even powers of the same Pauli blocks thus allowing reduction of the resulting multinomial terms in powers of the eigenvalues in the relevant multi-index space. This superposition of words is in direct analogy with the idea of the Feynman checkerboard [24],[25] and shows a link with the choice of a spinorial wavefunction in the first place which makes the introduction of a multi-index space inevitable due to the mixing of up and down terms.

5. **Possible analog machine implementations**

At this point we may start progressively searching for a class of analog machines capable of realizing the dynamics contained in the update equations of the previous section which should converge to a single signal analog machine incorporating unitary dynamics. The methodology introduced signifies the possibility of an embedding of discrete/symbolic structures in a continuum which serves as a "carrier" for the information of the encoded discrete structure. This is not necessarily restricted to the particular class of automata analyzed here and it may also be used in other paradigms like the so called, "natural computing" where algebraic manipulation of embedded symbolic structures can be achieved by means of analog filters and dynamical systems. Hence, the apparently insurmountable barrier between discrete/symbolic and continuous/analog representations can in some case be raised allowing a compromise for coexistence of the two.

Given an initial condition for both $\Phi$ and $\Xi$ vectors, we notice that apart from their amplitude only their relative phases are important. This situation is similar and can be mapped directly into a set of harmonic signals of the same frequency where the frequency is an independent parameter serving only as the "carrier" medium. If a clock signal is available then the relative phases can be recovered at any time for an update including their amplitudes. Using a multiplexing scheme with local connections, the equivalent of the kernel matrix can be imitated with appropriate weights in exact analogy with the way neural circuitry works, the only difference being the absence of a final nonlinear suppressor function as used in standard neural net applications.

To give an example, we also provide an interpretation of the diagonal propagator (15) as a transmission line model. Indeed, by flipping the up and down 2-vectors $[\xi_k, \phi_k]$ we cause a column wise mirror inversion of the **M** matrix which can now be put into direct correspondence with the transfer matrix of a classical quadruple circuit [32], [33] with a characteristic impedance $Z = Z_0/Z_L$ where $Z_L$ the load impedance corresponding to a reflection coefficient $G = (1+Z)/(1-Z)$ and a transmission coefficient $\gamma$ in the standard generic form coming from general solutions of the Telegrapher's equation

$$\mathbf{M}_k \cong \begin{bmatrix} \cosh(\gamma_k) & Z_k \sinh(\gamma_k) \\ Z_k^{-1} \sinh(\gamma_k) & \cosh(\gamma_k) \end{bmatrix} \qquad (16)$$

The form of (16) suggests to the immediate identifications

$$\lambda_{1,k} = Z \sinh(\gamma_k), \lambda_{2,k} = \cosh(\gamma_k), \lambda_{3,k} = Z^{-1} \sinh(\gamma_k)$$
(17)

The identification would be complete only if it satisfied the constraints

$$(\lambda_{1,k}/Z)^2 - \lambda_{2,k}^2 = (\lambda_{3,k}/Z)^2 - \lambda_{2,k}^2 = 1 \qquad (17)$$

As this is not possible directly, we can resort in the linearity of transmission elements and introduce two types of a "*Linear Transmission Line Network*" (*LTLN*). The first is given as a superposition of two quadruple circuits as shown in figure 2. We now have four degrees of freedom from the two transmission coefficients as well as their input impedances. Given the original formulation of the modified eigenvalues from (13b) it is possible to fully establish a correspondence with four unknowns in the resulting nonlinear identification equations. A full solution is provided in App. D while in figure 3(a-b) we show how the transmission coefficients vary across the original UCA arrays. As the strong nonlinearities that are present may cause certain engineering difficulties for shaping the transmission coefficients, we can also resort into a "Weighted LTLN" (WLTLN) method with a single quadruple and a multiplexing matrix as also detailed at the end of App. D. Indicative results are shown in figure 4 for the first matrix element. While randomly dispersed, these coefficients can be interpreted as just additional input impedances which are easier to construct. We should stress that this technique is fairly general and can be used to emulate any arbitrary 2 x 2 matrix operator. It is also immediately transferable in any optical media which can be described by an analogous transfer matrix formalism [34], [35]. Larger operators that are amenable to Kronecker factorization can also be emulated via multiplexing techniques.

While such a machine appears fit for a linear network structure like the one presented by (8) it does not yet utilize the full power of the circulant convolution filter build by construction in the kernel matrices. To do so one can make use of (12) or (13) and reverse the interpretation of the signals. Assume then a set of harmonics spanning an arbitrary spectral range and interpret any current configuration $\{\Phi_n(\omega_i), \Xi_n(\omega_i)\}$ as the complex spectrum coefficients of a single noisy signal which is neither 'even' nor 'odd'. The total power spectrum given from the amplitudes as $\sum |R_i(\omega)|^2$ should remain constant. From the general convolution theorem we know that any such operation can be inverted to a linear multiplication in the conjugated domain, therefore one may always interpret the continuous analog of the previous section dynamics as a deconvolution of the collective signal in the time domain. Additionally, the exchange between frequency and wavelength domains does not affect the above observation but switches the implementation to an appropriate linear spatial medium.

6. **Discussion and conclusions**

We have reported some new results on the analysis of UCA with the purpose of finding efficient implementations with analog machines operating on continuous signals. This was supported by a previously introduced technique where multi-dimensional lattices are unfolded and projected on one dimensional arrays, such that they are amenable to certain techniques from signal theory utilizing circulant convolution filters based on DFT modules. Given the linearity of all such automata, their diagonal representation on a Fourier basis provides a natural framework for basing the construction of analog or optical machines either on the frequency or the wavelength domain, with the continuous substrate being the information carrier. This opens a new way of interpreting discrete/symbolic information encoding into continuous media which partially raises the absolute distinction between these two domains without sacrificing anything from the original discrete formulation. On the other hand, it opens an area of possible new questions as for instance the case of the principle of indistinguishability between data and code description as is known to hold in classical $\lambda$-Calculus and string rewriting automata of classical computational theory. In the case of analog machines encoding symbolic information it is also possible to ask for cases of indistinguishability between code and machine description leading to certain classes of self-modifying automata. Whether this is also compatible with unitarity is under investigation and it could be used as a means to construct toy models for shedding some light in the hard problems of incompatibilities between gravitational theories and unitary quantum mechanics.

As a last step, we would like to discuss the special role played by the circulant kernels introduced in this work as the sole means for effectively storing topological information of any discrete lattice and preserving correlations of the equivalent dynamics. An immediate consequence of this is that one can increase or decrease the effective dimensionality by simply altering the kernel structure. For instance, one could use an unbounded dimensionality for a single hyper-cubic neighbourhood to store the same information from a lattice $L^D$ into a lattice of a single hyprecubic neighbourhood of $L^D$ dimensions. As a result, one would form the circulant kernel as an $L^D$ Kronecker product $\mathbf{C}_1 \otimes \mathbf{C}_2 \otimes ... \otimes \mathbf{C}_{L^D}$ over a collection of circulant sub-blocks. Such a procedure naturally produces certain self-similar kernels of which the structure is always an arithmetic fractal as can be verified by a direct computation.

This particular duality that allows exchanging between many states of a D-dimensional cell lattice with many subsets of neighbors of a single hypercubic cell neighborhood leads to an interesting comparison with the case of entangled states. For this to make sense, we also need to change the encoding of the initial conditions in order to reflect the realization of all the combinatoric powerset of a given number of states. While the correspondence for the kernel structure has been shown in App. C, construction of such initial conditions will have to follow the same recursive structure as used in the construction of the sub-kernel characteristic rows. As the sole role of the different kernel structure is to alter the resulting eigenvalue spectrum, the diagonal representation of the resulting dynamics will still be the same as in the original form given here albeit not necessarily as that prescribed in the original Bialinycki-Birula's paper. Any such machine then should be realizable by one of the methods of section 5. This theme is under investigation and results will be reported in a subsequent publication.

**Appendix A.**

The standard way of expressing the integers in various alphabets via the polynomial representations is well known. In well formed alphabets, this is always given via powers of the alphabet base hence we may call these representations as "harmonic" or "ordered" to distinguish them for others. There are also certain other irregular representations like the Fibonnaci representation based on the Zeckendorf theorem [27] where the alphabet basis is not finite and is given by an infinite sequence with or without a composition law. Any such sequence is called complete if it has the expressive power to represent all integers with binary digits. A sufficient but not necessary condition for completeness has been given via the partial sums of an arbitrary integer sequence [28] We shall next prove by an inductive argument that the partial products of any positive definite sequence also satisfy a generalized completeness in higher symbolic alphabets with digits > {0, 1}.

A well known method for addressing the values of multi-dimensional arrays in a linear one dimensional arrangement which could also be the Random Access Memory of a computing machine is given by the following scheme. Given a set of dimensions of the array which is to be defined say as $A(D_1, D_2, D_3)$ over a linear list of data of length $L = D_1 D_2 D_3$, one defines a set of three pointers $\{i, j, k\} \in [0,...,D_i]$ and establishes a unique correspondence with a pointer $n$ to the linear list as

$$n = i + (j-1)D_1 + (k-1)D_1 D_2 \tag{A.1}$$

Assume then that the A array holds only the values of the pointer $n$ running on all integers in the interval $[0,...,D_1 D_2 D_3]$. Then [A1] is a perfect representation of all integers in that interval. Assume also that we increase the dimensions of the array by one as say, $A(D_1, D_2, D_3, D_4)$ pointing to a list of new length given by the product of all four dimensions. We should still be able to write

$$n = i + (j-1)D_1 + (k-1)D_1 D_2 + (m-1)D_1 D_2 D_3 \tag{A.2}$$

By induction one can expand the above to any infinite sequence of positive numbers serving as the dimensions of an infinite dimensional array thus establishing a correspondence of all integers with the sequence of partial products $\prod_{i=1} D_i$ for any strictly positive sequence of integers $D_1, D_2, …$

**Appendix B.**

According to the prescription of section **4**, the dissected lattice contains two identical copies with appropriately separated subsets of complex values from any initial condition. Alternating neighbors correspond to adjacent cubic blocks of 8 sites while each update move proceeds with gliding over these blocks. It is then straightforward to write down the exact form of the defining vectors for each circulant kernel of equations (12a-b) as connectivity matrices for the correct addressing of the original sites in the unfolded form using the same prescription of the special index arithmetics shown in App. A. Specifically, each update rule uses half of the 8 sites in the reduced lattices block for spin-up and spin-down elements so that we have

$$\begin{aligned}
\mathbf{c}_1 &= k[\alpha, \beta, 0, ...0; \beta, \alpha, 0...0] \\
\mathbf{c}_2 &= k[0, ..., 0; -\beta, \alpha 0, ..., 0, -\alpha, \beta, 0...0] \\
\mathbf{c}_3 &= k[-\alpha, \beta, 0, ..., 0; -\beta, \alpha, 0...0] \\
\mathbf{c}_4 &= k[0, ..., 0; \beta, \alpha, 0, ..., 0, \alpha, \beta, 0...0]
\end{aligned} \tag{B.1}$$

In (B1) we have kept only the different phases with their common radius given by $k = 1/2\sqrt{2}$. The ";" is used to denote the end of the first row in $c_{1,3}$ or the first plane for $c_{2,4}$ in the unfolded form of the original 3D lattice. All characteristic rows constructed this way are mutually orthogonal satisfying $(\mathbf{c}_i / 2k)(\mathbf{c}_j / 2k) = \delta_{ij}$ The addressing scheme used maintains zeros in specified intervals of lengths [3:$N$] and [$N$+3:$N^3$] for $c_{1,3}$ corresponding to the lower face of the first cubic block while for the second upper face one needs three zero intervals as [1:$N^2$], [$N^2$+3: $N^2$+$N$] and [$N^2$+$N$+3:$N^3$]. For periodic boundary conditions, each vector will have to be shifted cyclically by one position to the left. The eigenvalue spectrum can be found analytically with the aid of the Vandermode matrix on the N roots of unity $\omega_n = \exp(2\pi i n/N)$ which is filtered out at exactly four positions and using $\alpha\beta$=1, $\alpha^2$ = $i$, $\beta^2$ = -$i$ and $k \to k/\sqrt{N}$ giving rise to the following finite Fourier polynomials

$$\lambda_i^1 = k\alpha[\mathbf{i} + \omega_i + (1 + \mathbf{i}\omega_i)\omega_i^N]$$
$$\lambda_i^2 = -k\alpha\omega_i^{N^2}[1 - \mathbf{i}\omega_i + (\mathbf{i} - \omega_i)\omega_i^N]$$
$$\lambda_i^3 = -k\alpha[\mathbf{i} - \omega_i - (1 + \mathbf{i}\omega_i)\omega_i^N] \quad \text{(B.2)}$$
$$\lambda_i^4 = k\alpha\omega_i^{N^2}[1 + \mathbf{i}\omega_i + (\mathbf{i} + \omega_i)\omega_i^N]$$

The easiest way to create a reduced dimensionality circulant kernel is the tensorial (Kronecker) product of $D$ block matrices with sparse bands of which the elements cover each side of a hypercubic neighborhood. A general decomposition of a random matrix into a Kronecker product is possible only if the original elements can be factorized in a particular way. In the case of a circulant kernel though, we have a serious reduction due to the fact that all rows are cyclic permutations of the first one. Hence, if we can factorize appropriately the first row, the problem is solved. In the case of UCA as defined by Birula, all rows will contain only four cases of coefficients which are of the form $\pm \exp(\pm \mathbf{i}\pi/4)/2\sqrt{2}$.

We then request for each of the resulting circulant kernels to have the form $\hat{C}_i = \otimes_{i=1}^3 \mathbf{C}_i$ with $\mathbf{C}_i$ being unique $N$ x $N$ circulant blocks. In order to find the block elements we need a recursive procedure which will only make use of the characteristic vectors $c_i$ of the first rows of $C_i$. The final $N^3$ row $c$ will then be interpreted as a successive product of length $N$ sub-vectors which should satisfy a recursive relation as

$$\mathbf{c}^{n+1} = \left[c_1^n \mathbf{c}^n, c_2^n \mathbf{c}^n, c_3^n \mathbf{c}^n, 0...0\right] \quad \text{(B.3)}$$

Equating this with each of the original rows in (B1) and taking logarithms will then lead to a linear system for the exponents. While such a process will require only $nD$ unknown parameters, the final row to be compared contains in general $D^n$ values hence only if these have a sufficient redundancy is factorization possible. In our case only four of them are non-zero for each of (B1) so that we can safely choose $c_1 = I$, and take the rest two rows with only a pair of unknown parameters that will give four combinations of products $c_i c_j$. Taking the radius as either $\sqrt{k}$ or $\mathbf{i}\sqrt{k}$ depending on the sign leaves four unknown phase parameters which can be taken as $\varphi_i = k_i \pi/4$. Factorization then becomes equivalent to the condition $k_i + k_j = \pm 1$ where $k_i$ rational numbers. Such a problem does not have a unique answer as it resolves to the geometric problem of finding appropriate subdivisions of arcs on the unit circle of which instances are countably infinite. Thus factorization is always possible for this class of kernels but at the cost of introducing an uncertainty.

**Appendix C:**

We prove the existence of a map from the set of Kronecker products of equal dimension blocks to the lexicographically ordered powersets of symbolic strings from an arbitrary alphabet base. To do so we need a special correspondence where the algebraic multiplication at each step of combining two $L \times L$ block matrices $A$ and $B$ is replaced with a concatenation operation as $a_{ij} b_{kl} \to [a_{ij}, b_{kl}]$. In a total of $D$ such blocks one obtains a set of $L^{2D}$ lists. Assume also a second step at which the elements of each list are replaced with the expansion of their indices in an alphabet of base $b$, the total map being then

$$a_{ij} b_{kl} \to [(i + n(j-1))_b, (k + n(l-1))_b] \tag{C.1}$$

The outcome of such a procedure is just a rearrangement of the $L^{2D}$ symbolic strings when $b = L$ for any $L$ prime otherwise taken *modulo* one of its factors or more of them. To give a concrete example, we may take the product $\mathbf{a} \otimes (\mathbf{b} \otimes \mathbf{c})$ where all *a*, *b* and *c* are 2 x 2 matrices. Then the result of applying [B1] will give exactly 64 binary strings which exhaust the powerset of all 6 bit strings. The correspondence of the element indices of the final matrix to the lexicographic ordering of any such powerset can be made exact with a special reading protocol which is known as a "*Morton Code*" [29], [30] and it is a discrete sampling of a *Lebesque Z-Curve,* a well known space-filling differentiable curve [31].

**Appendix D.**

From the superposition of quadrupoles in figure 2, and an input vector $\mathbf{X} = [\xi, \varphi]$ we derive the identification equations from

$$(\mathbf{T}_1 + \mathbf{T}_2) \cdot \mathbf{X} = \mathbf{M} \cdot \mathbf{X} \tag{D.1}$$

Using the explicit form of the eigenvalues from the original circulant kernels we derive the equivalence relations

$$\begin{aligned} \cosh(\gamma_1) + \cosh(\gamma_2) &= \lambda_1 \lambda_3 + \lambda_3 \lambda_4 \\ Z_1 \sinh(\gamma_1) + Z_2 \sinh(\gamma_2) &= \lambda_1^2 + \lambda_3^2 \\ Z_1^{-1} \sinh(\gamma_1) + Z_2^{-1} \sinh(\gamma_2) &= \lambda_3^2 + \lambda_4^2 \end{aligned} \tag{D.2}$$

The first two are immediately decoupled to obtain the transmission coefficients as

$$\gamma_1 = \cosh^{-1}(\lambda_1 \lambda_3), \quad \gamma_2 = \cosh^{-1}(\lambda_3 \lambda_4) \tag{D.3}$$

The remaining nonlinear system takes the simplified form

$$Z_1 + Z'_2 = C_1, \quad \frac{A}{Z_1} + \frac{B}{Z'_2} = C_2 \tag{D.4}$$

In (D4) we have made the substitutions

$$Z'_2 = Z_2 \sinh(\gamma_2)/\sinh(\gamma_1) = Z_2 \sqrt{A/B}$$
$$A = (\lambda_3 \lambda_4)^2 - 1, \quad B = (\lambda_1 \lambda_3)^2 - 1 \tag{D.5}$$
$$C_1 = (\lambda_1^2 + \lambda_3^2)/\sqrt{A}, \quad C_2 = (\lambda_3^2 + \lambda_4^2)\sqrt{B}$$

Solutions of (D4) are given as $C_1 - \rho_1, \rho_2$ where $\rho_i$ the roots of the trinomial

$$C_2 X^2 + (A - B - C_1 C_2) X + B C_1. \tag{D.6}$$

The simpler case of the weighted LTLN has the identification equations

$$\mathbf{T} \cdot \mathbf{W} \cdot \mathbf{X} = \mathbf{M} \cdot \mathbf{X} \tag{D.7}$$

This is then directly invertible as $\mathbf{W} = \mathbf{T}^{-1} \cdot \mathbf{M}$ in terms of a prescribed transmission coefficient and impedance. Inversion of $\mathbf{T}$ is just a sign inversion in the anti-diagonal. From (D1) and (D7) we also have the intermediate identity

$$\mathbf{T}_1 + \mathbf{T}_2 = \mathbf{T} \cdot \mathbf{W} \tag{D.7}$$

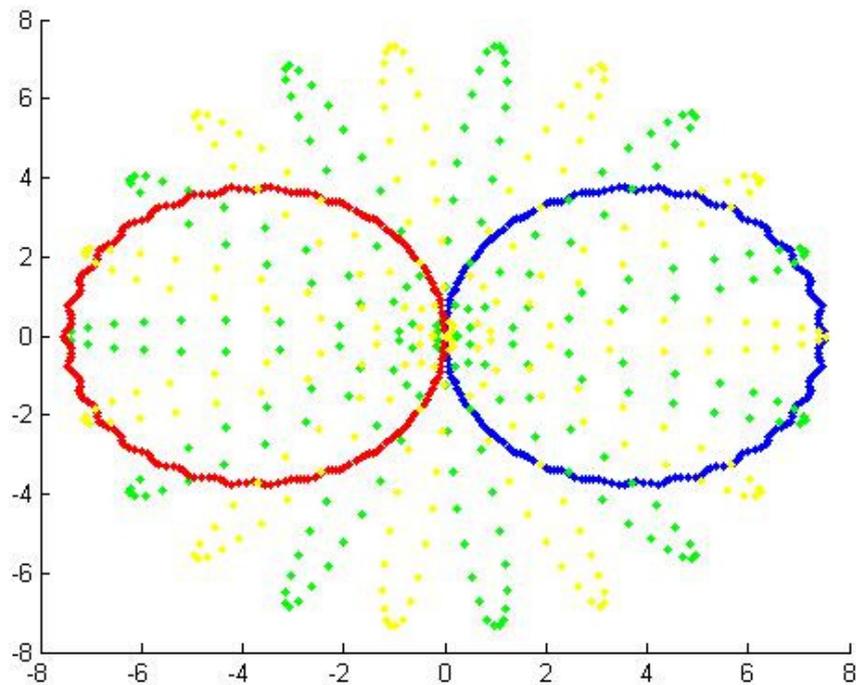

**Fig. 1**a. Eigenvalue spectra from FFT of kernel rows: (a) for the original circulant kernels of (12a-b).

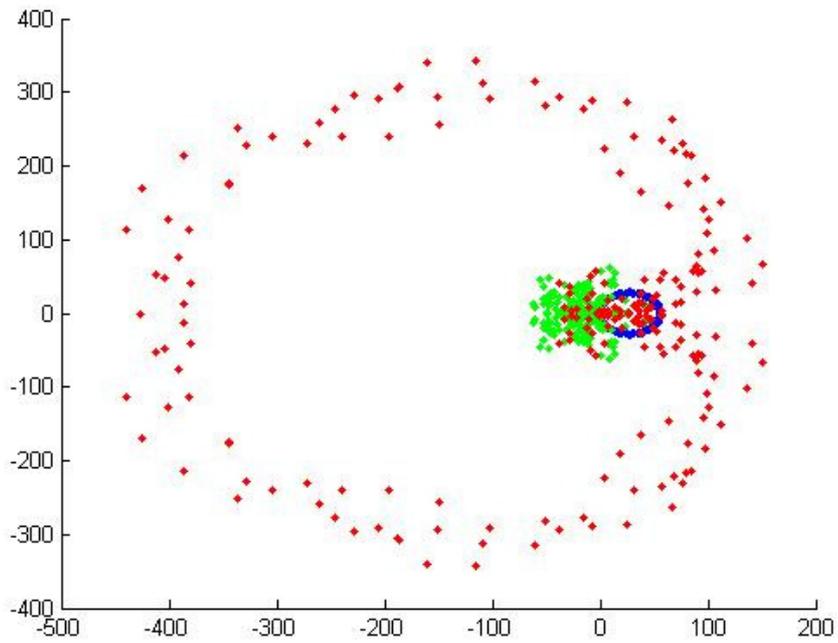

**Fig 1**b. Eigenvalue spectra from FFT of kernel rows for the three composite kernels of (13a-b).

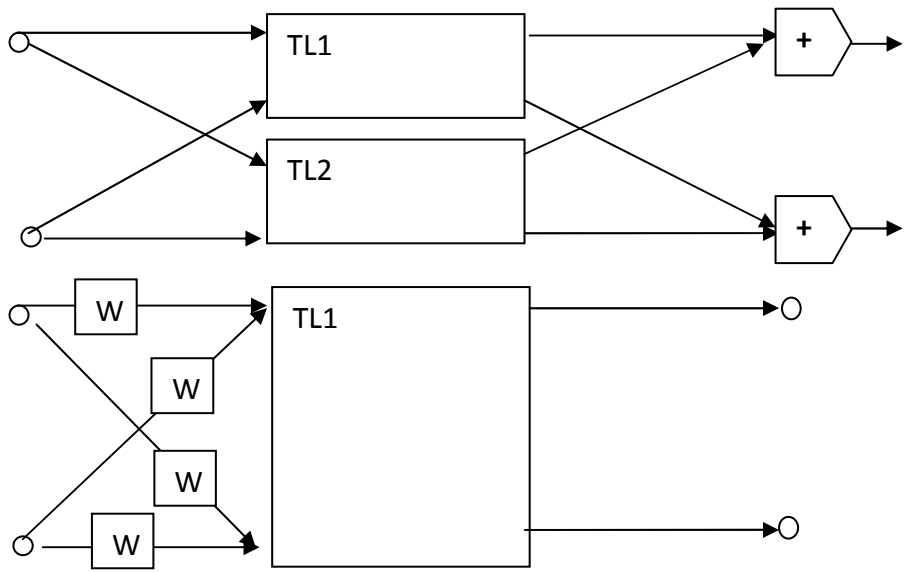

Figure 2, (a) Schematic for a double quadrupole TLN and, (b) the same for a single weighted TLN

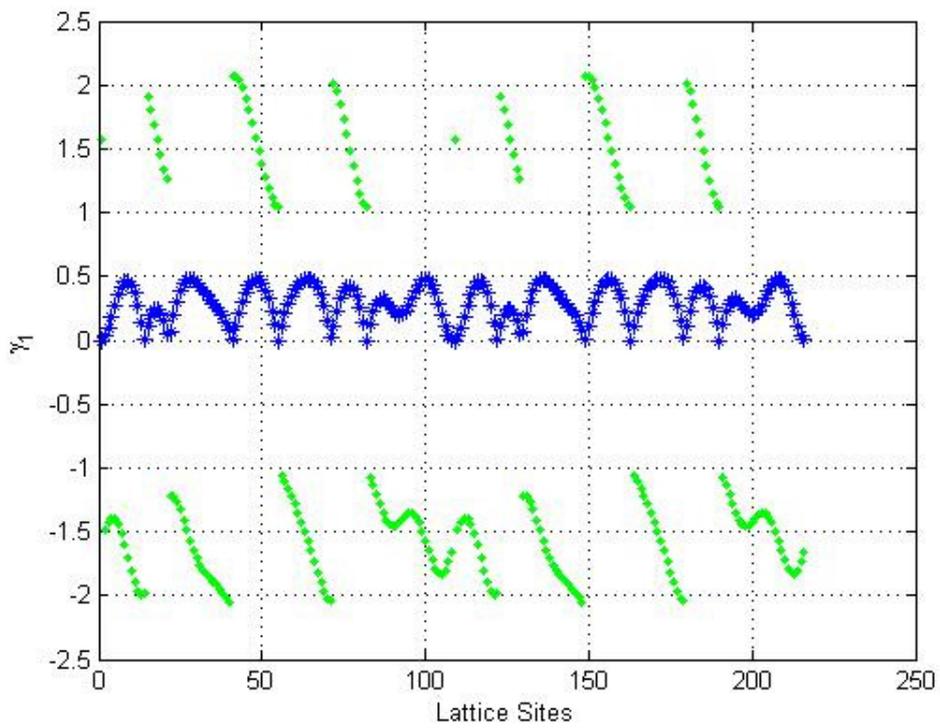

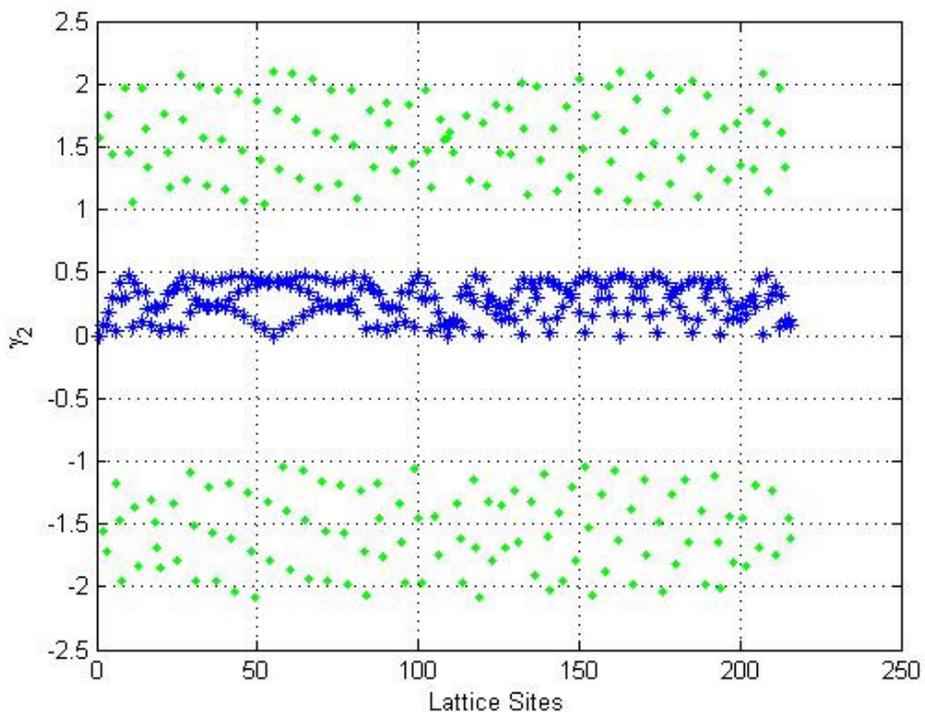

**Fig 3**, Variation of the TLN transmission coefficients (Real: (*), Imag (.)).

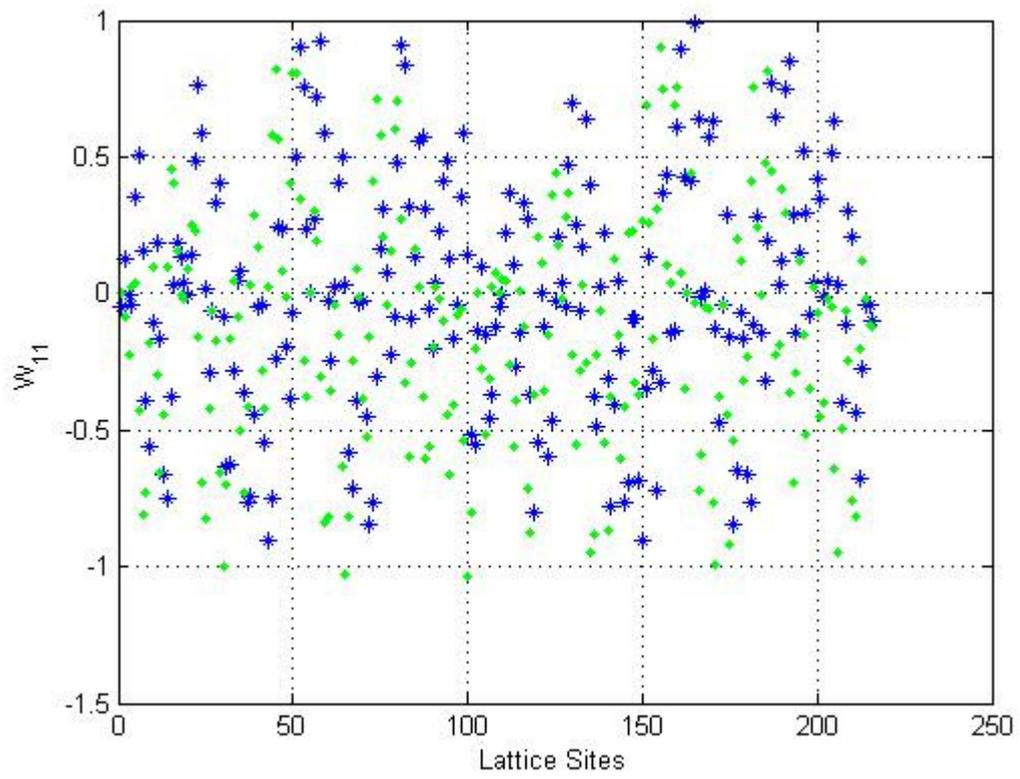

**Fig. 4**, Indicative variation of the first of WTLN coefficients (Real: (*), Imag (.)).